\documentclass[journal]{IEEEtran}

\usepackage{amsmath,amssymb}
\usepackage{booktabs}
\usepackage{array}
\usepackage{cite}
\usepackage{graphicx}
\usepackage{url}
\graphicspath{{generated/}}

\newcommand{\WidthTrain}{\texttt{WIDTH\_TRAIN}}
\newcommand{\PhaseTrain}{\texttt{PHASE\_TRAIN}}

\newcommand{\ModeZero}{M0}
\newcommand{\ModeOne}{M1}
\newcommand{\WidthOne}{W1 (3/5 cycles)}
\newcommand{\WidthThree}{W3 (4/7 cycles)}
\newcommand{\CompletePSR}{\ensuremath{\mathrm{PSR}_{\mathrm{complete}}}}
\newcommand{\BasePSR}{\ensuremath{\mathrm{PSR}_{\mathrm{base}}}}
\newcommand{\ServicePSR}{\ensuremath{\mathrm{PSR}_{\mathrm{service}}}}

\newcommand{\ResultConfigHash}{\texttt{593f2287c9b49ead}}
\newcommand{\ResultPacketsPerCell}{2000}

\newcommand{\ResultOpsPerSymbol}{320}
\newcommand{\ResultStateBytes}{1024}
\newcommand{\ResultRendererRate}{256}

\title{An Implant-to-Wearable IR-UWB Transmitter--Receiver Architecture and Layered Protocol for High-Density Brain--Computer Interfaces}

\author{G.~D.~Su\textsuperscript{*} and Y.~Mo%
\thanks{G. D. Su is an Associate Professor with Hangzhou Dianzi University, Hangzhou, China.}%
\thanks{Y. Mo is an independent researcher and is with BroadLink Co., Ltd.}%
\thanks{\textsuperscript{*}Corresponding author: G. D. Su (e-mail: guodong@hdu.edu.cn).}}

\begin{document}

\maketitle

\begin{abstract}
High-rate neural telemetry requires an implant-to-wearable link whose circuit partition, waveform mapping, packet processing, and failure semantics are mutually consistent.
This paper specifies a chip-oriented, one-way impulse-radio ultra-wideband (IR-UWB) uplink comprising an implanted packet engine and free-running transmitter, a tissue-proxy channel, and a wearable mixed-signal receiver and digital baseband.
The implant path maps coded packets to \ModeZero{} pulse-width modulation (PWM) or layered \ModeOne{} PWM plus differential binary phase-shift keying (DBPSK); a nominal 300-MHz bi-phase ring oscillator (RO) supplies selected phases to the edge combiner, pulse-control, power-amplifier, and implant-antenna path.
The RO/edge-combining transmitter circuit is attributed to Lei \emph{et al.} (2024) and is not claimed as an original circuit contribution.
At the wearable, an antenna and analog front end feed one envelope branch for acquisition, timing, and PWM soft information and one time-shared burst-I/Q branch for optional DBPSK demodulation.
A known \WidthTrain{} field updates short- and long-pulse templates online, produces log-likelihood ratios (LLRs), and selects a conservative packet-local fallback when training is unreliable.
The deterministic encode/decode chain specifies robust Header protection and independent Base/Enhancement serialization, FEC, interleaving, CRC, and AEAD verification, so Enhancement failure can yield explicitly labeled Base-only service without being counted as complete delivery.
Evaluation uses an event-timed, sample-domain channel/front-end model with RO variation, fractional-delay multipath, filtering, gain control, quantization, and clipping, followed by a frozen 80{,}000-packet held-out campaign, a paired 256/512-sample DBPSK sensitivity check, and a separate benign coding study.
The frozen held-out campaign provides complete packet-level waterfall and predeclared population-shift evidence rather than relying only on uncoded BER.
The paired check detects material phase-path sampling sensitivity near the
waterfall and therefore does not establish convergence of the frozen
256-sample implementation.
In the separately declared benign single-tap sensitivity condition, the normal single-copy M0 and M1 profiles each observed 500/500 successful packets at 18\,dB, while the fixed two-copy soft-combining profile observed 500/500 for both modes at 15\,dB.
Together, these results demonstrate end-to-end architectural and protocol feasibility, quantify the gain available from robust coding/combining profiles, and turn the remaining receiver and channel margins into concrete priorities for the next implementation stage.
This pre-silicon study provides an auditable path toward receiver-chip implementation and measured biological-channel validation.
\end{abstract}


\section{Introduction}
\label{sec:introduction}

High-channel-count neural recording can produce sustained data rates that are difficult to export under implant power, area, and temperature constraints \cite{Steinmetz2021}.
Impulse-radio ultra-wideband (IR-UWB) telemetry is attractive in this setting because a duty-cycled implanted transmitter can generate short RF bursts while moving accurate timing, data conversion, and most digital processing to an external unit \cite{Eskandari2024}.
Recent silicon demonstrates substantial progress in high-order crystal-less transceivers, hybrid impulse modulation, multi-rate and data-to-time transmitters, and noncoherent receivers \cite{Song2022,Lei2026Transceiver,Ding2026,Hayati2025,Eskandari2025Receiver}.
Event-based packet encoding and packet-level frequency control have also been reported for compressive neural telemetry \cite{Liu2026}.
Lei \emph{et al.} further report a crystal-less IR-UWB transceiver with an implant-side receiver and wireless clock calibration \cite{Lei2026Transceiver}.
Accordingly, neither complete transceiver capability, implant reception, nor packetization is absent from prior art.
Public work typically emphasizes an individual transmitter, receiver, transceiver circuit, or laboratory link.
Within our bounded review, we did not identify this exact combined scope: a chip-oriented implant-to-wearable partition joined to a deterministic two-lane processing chain, packet-local receive adaptation and fallback, and a frozen sample-domain held-out evaluation.
This statement describes the review boundary, not an exhaustive literature, product, standards, or patent novelty search.

\begin{figure*}[t]
  \centering
  \includegraphics[width=0.98\textwidth]{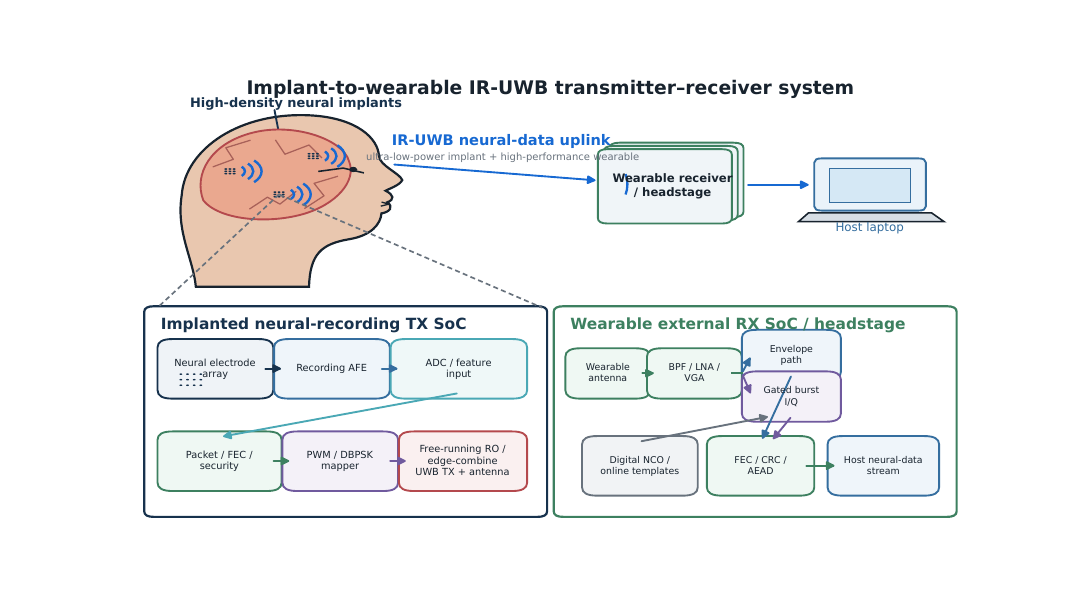}
  \caption{System-level application context for the
  implant-to-wearable IR-UWB transmitter--receiver architecture. Multiple
  high-density neural interfaces feed an implanted recording and telemetry
  SoC; the wearable receiver recovers the layered neural-data stream and
  forwards it to a host. The use case follows recent neural IR-UWB research
  \cite{Lei2024,Ding2026}; the artwork and technical partition are original.}
  \label{fig:bci-application-context}
\end{figure*}

The distinction is one of scope rather than a claim that the individual blocks
are unprecedented.
Transmitter-centered neural IR-UWB publications primarily optimize pulse
generation, modulation order, radiated link, and implant energy
\cite{Lei2024,Ding2026,Hayati2025,Liu2026}; receiver-centered work establishes
specific demodulator circuits and sensitivity/energy tradeoffs
\cite{Eskandari2025Receiver}; and recent transceiver silicon adds an implant
receive path and wireless clock calibration \cite{Lei2026Transceiver}.
This paper instead freezes the interfaces among an attributed implant
transmitter, a proposed wearable receiver, matching modulation/demodulation,
packet coding and verification, channel/front-end observations, and explicit
service outcomes.
It should therefore be judged as a pre-silicon system/protocol and receiver
architecture study, not as a replacement for those measured circuit results.

This study starts from the crystal-less all-digital IR-UWB transmitter reported by Lei \emph{et al.} \cite{Lei2024}.
In particular, its nominal 300-MHz free-running bi-phase ring oscillator (RO), selected-phase edge combining, and pulse-generation organization are attributed to that work.
We do not claim that transmitter circuit as original.
Our implant-side work is the architectural connection from neural-data packetization and independent lane coding to a bounded \ModeZero{}/\ModeOne{} mapper and to the attributed pulse generator.
The question is therefore end-to-end: how should an implant and wearable be partitioned, and how should their encoding and decoding chains correspond, when the transmitted timebase is free-running, the received pulse shape varies, and the optional phase lane can fail?

The central design choice is asymmetric complexity.
The implant retains the packet engine, lane mapper, free-running RO/edge combiner, pulse-width and polarity controls, power amplifier (PA), and antenna.
Across the tissue channel, the wearable contains an antenna and analog front end (AFE), one envelope path, one time-shared burst-I/Q path, a digital baseband, and a host interface rather than a continuously sampled multi-GS/s waveform digitizer.
The envelope path supports acquisition, timing observations, and pulse-width LLRs for a robust Base lane.
The I/Q path produces one complex observation per selected burst for an optional DBPSK Enhancement lane.
This organization is chip-oriented in the limited architectural sense that it exposes functional blocks, timing windows, state, arithmetic, power-gating boundaries, and dataflow on both sides of the channel; it does not claim a fabricated implementation, area, power, or timing closure.
Figure~\ref{fig:bci-application-context} summarizes the intended BCI use case
and the end-to-end hardware architecture.

The paper makes five bounded contributions:
\begin{enumerate}
    \item an end-to-end, chip-oriented partition from the implant neural-data and packet path through the attributed RO/edge-combining transmitter, tissue channel, and wearable AFE/baseband/host path;
    \item a paired modulation/demodulation architecture in which \ModeZero{} maps Base data to PWM and \ModeOne{} layers DBPSK Enhancement transitions over the same PWM bursts, with envelope and gated burst-I/Q receive branches;
    \item a deterministic Header and independent Base/Enhancement encode/decode chain specifying processing order, FEC, integrity checks, resets, and delivery semantics, while leaving production interoperability parameters to a future profile;
    \item packet-local online width templates, quality tests, and deterministic fallback that preserve verified Base service while distinguishing it from complete \ModeOne{} delivery; and
    \item a frozen channel- and sample-domain evaluation using common front-end observations, held-out device/channel clusters, coding comparisons, and separately declared benign sensitivity conditions.
\end{enumerate}

The numerical evidence uses common front-end observations, frozen
training/validation decisions, two independently seeded synthetic held-out
populations, and cluster-aware uncertainty intervals.
Earlier small-sample proxy studies are not copied into the reported result
claims.

The remainder of the paper defines the system boundary, implant transmitter, wearable receiver, protocol,
training policy, and evaluation method before presenting the frozen
simulation results and publication limitations.

\section{System Model and Scope}
\label{sec:system}

\subsection{Asymmetric Uplink}

The modeled system is a one-way neural-data uplink with three physical partitions.
On the implant, a neural-data interface feeds buffering and a packet engine, Header and lane encoders, an \ModeZero{}/\ModeOne{} symbol mapper, a free-running RO and selected-phase edge combiner, pulse-width and polarity controls, a PA, and an implant antenna.
The radiated bursts traverse the package--antenna--tissue channel.
At the wearable, an external antenna and AFE provide filtering, gain control, and sampled envelope and gated I/Q observations; a digital baseband performs acquisition, timing, demapping, deinterleaving, FEC decoding, integrity/authentication checks, fallback, and host delivery.
The host interface exports verified data and explicit status rather than unverified demodulator output.

This transmitter--receiver terminology names the two endpoints of the uplink.
The manuscript does not define a bidirectional transceiver product: there is no implant receiver, external downlink transmitter, acknowledgement path, or over-the-air transmitter-control protocol in the technical scope.
Reference-assisted transmitter calibration, an implant receiver, and an external downlink remain nontechnical out-of-scope interfaces only.

\begin{figure*}[t]
  \centering
  \includegraphics[width=0.98\textwidth]{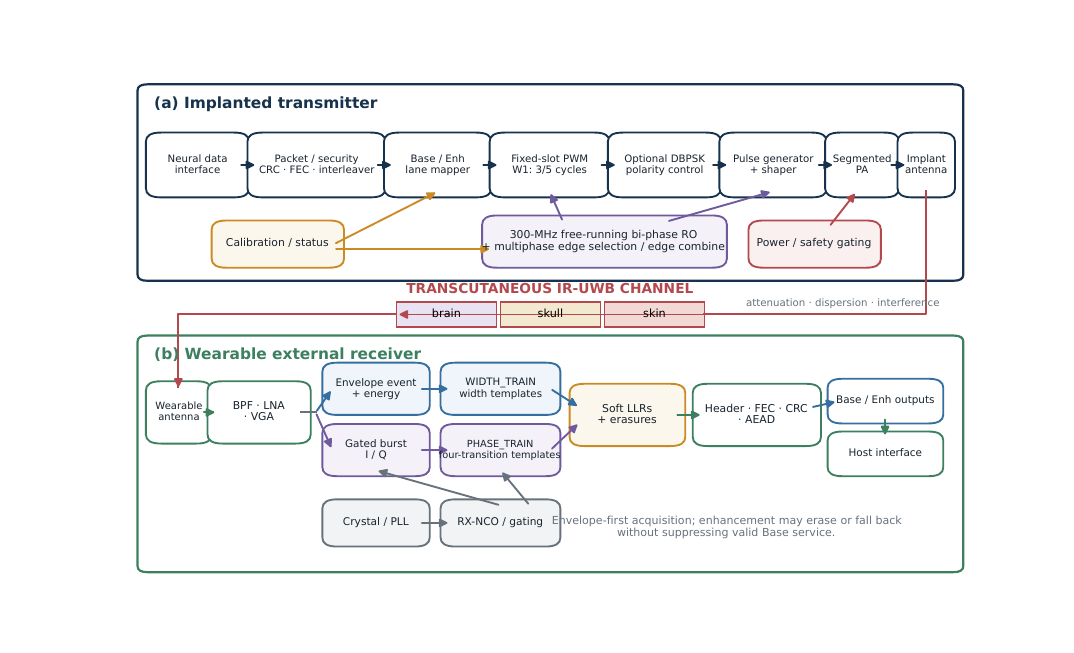}
  \caption{End-to-end chip-oriented partition of the one-way
  implant-to-wearable uplink. The implant RO, selected-phase edge-combining,
  pulse-generation, and PA organization is adopted from Lei \emph{et al.}
  \cite{Lei2024}; the packet/lane interfaces, wearable receiver partition,
  decoder flow, and cross-layer service semantics are specified in this work.
  W1 3/5-cycle PWM is shown as the baseline programmable width profile.
  The tissue blocks denote a synthetic channel boundary, not an anatomical or
  measured propagation model.}
  \label{fig:end-to-end-architecture}
\end{figure*}

The implant emits fixed-slot bursts whose RF-cycle count carries a pulse-width bit.
In \ModeZero{}, pulse width alone carries one coded Base bit per symbol.
In layered \ModeOne{}, the same pulse-width bit carries Base data and an optional $180^{\circ}$ differential polarity transition carries one coded Enhancement bit.
The latter does not require an absolute transmitter phase reference; it does require a valid differential observation at the external receiver.
Preamble, training, SFD, Header, payload, and pilots all pass through the same physical transmitter and receiver paths; the receiver is not granted a separate ideal observation for control fields.

The free-running transmitter RO introduces a common timebase uncertainty into symbol period, burst width, and RF carrier offset.
For symbol index $k$, an abstract correlated model is
\begin{equation}
 f_{\mathrm{RO}}[k]
 = f_0\!\left(1+\epsilon_{\mathrm{dev}}+\epsilon_{\mathrm{slow}}[k]
 +\epsilon_{\mathrm{rw}}[k]+\epsilon_{\mathrm{burst}}[k]\right),
\end{equation}
where $f_0$ is nominal RO frequency and the dimensionless $\epsilon$ terms represent device offset, slow drift, random walk, and burst-correlated perturbation.
For the frozen campaign, define the ppm-domain components
$\widetilde\epsilon_\bullet=10^6\epsilon_\bullet$; the implemented model is
\begin{align}
 \widetilde\epsilon_{\mathrm{dev}}&=A_{\mathrm{dev}},\\
 \widetilde\epsilon_{\mathrm{slow}}[k]
 &=A_{\mathrm{start}}e^{-(k+96)/48}
   +A_{\mathrm{end}}\frac{k}{K-1},\\
 \widetilde\epsilon_{\mathrm{rw}}[k]&=\sum_{i=0}^{k}\eta_i,\\
 \widetilde\epsilon_{\mathrm{burst}}[k]&=120(w_k-\bar w),
 \label{eq:ro-statistical-model}
\end{align}
where $K$ is packet length, all numerical values are in ppm,
$A_{\mathrm{dev}}\sim\mathcal N(0,6000^2)$,
$A_{\mathrm{start}}\sim\mathcal N(0,2500^2)$,
$A_{\mathrm{end}}\sim\mathcal N(0,220^2)$, and
$\eta_i\overset{\mathrm{iid}}{\sim}\mathcal N(0,2.5^2)$.
The same RO error controls symbol period, the 4.2-GHz burst carrier,
burst width, carrier offset, and accumulated phase within a packet.
Independent zero-mean Gaussian draws add 4-ps RMS symbol-boundary jitter,
an 8-ps packet-shared width offset plus 2-ps per-symbol width jitter, and a
2-ppm receiver-LO offset.
Despite the ``device'' label, the current implementation redraws these
packet-level RO variables for every packet; it does not preserve one device
offset across repeated packets.
The wearable receiver is not given the true realization of these terms.
It estimates the transmitted time axis from preamble observations and tracks it with a local numerically controlled oscillator (NCO).
Its precise local reference assists reception only; it neither locks nor retroactively corrects the radiated implant waveform.

\subsection{Channel and Front-End Observation}

The publication campaign renders a complex-envelope burst train using the actual event times generated by the RO model.
The received waveform is represented as
\begin{equation}
 r(t)=\sum_{k}\sum_{m} a_m x_k(t-t_k-\tau_m)+n(t),
 \label{eq:channel}
\end{equation}
where $x_k(t)$ is the transmitted short or long burst, $t_k$ is its event time, $a_m$ and $\tau_m$ are complex channel gain and delay, and $n(t)$ is equivalent front-end noise.
Fractional delays, causal front-end filtering, gain control, quantization, and clipping are applied before detector-specific processing.
The SNR reference plane is the noiseless channel output immediately before
addition of $n(t)$ and before AFE filtering.
For each packet, the renderer defines
\begin{equation}
 \mathrm{SNR}_{\mathrm{eff}}=10\log_{10}(E_{\mathrm{ref}}/N_0),
 \label{eq:snr-reference}
\end{equation}
where $E_{\mathrm{ref}}$ is the channel-gain-scaled energy of one isolated
four-cycle reference burst and $N_0$ is the complex white-noise energy
spectral density used by the sample-domain renderer.
The nominal sweep value is adjusted by the sampled path-loss deviation,
diversity term, and blocker penalty before this noise level is set.
It is therefore a controlled simulation reference, not transmitter power,
RSSI, sensitivity, range, or $E_b/N_0$.
No conversion from the reported nominal-SNR axis to a link budget, implant
transmit power, communication range, or receiver sensitivity is valid without
measured and jointly calibrated PA, antenna, package, biological-channel, AFE,
and noise data.

Three channel populations are separated by purpose: an AWGN sanity set, an in-distribution statistical tissue-proxy population, and a shifted hold-out population.
Their parameterization is informed by published implant-channel modeling
\cite{Bahrami2015}, but the proxy populations are not measurements of a
person or animal.
The tissue-proxy populations are hierarchical: cluster-level path loss and
delay are shared by eight generated proxies, proxy-level perturbations recur
with a proxy, and packet-level path-loss, blocker, diversity, tap, phase, and
noise draws are resampled.
Device/channel cluster identifiers remain attached to packets so uncertainty can be computed at the cluster level rather than by treating all packets as independent.
The frozen numerical values and sampling hierarchy are summarized in
Sec.~\ref{sec:reproducibility}.

\subsection{Service Outcomes}

The receiver records a staged outcome for every attempted packet:
acquisition, SFD, Header, Base, Enhancement, complete delivery, and Base-only service.
For \ModeOne{},
\begin{align}
\CompletePSR
 &= \Pr(H\cap B\cap E),\\
\BasePSR
 &= \Pr(H\cap B),\\
\ServicePSR
 &= \Pr(H\cap B),
\end{align}
where $H$, $B$, and $E$ denote verified Header, Base, and Enhancement events.
Although the last two probabilities share an event definition in the present one-packet receiver, ``service'' is reported separately to expose the policy decision that the verified Base PDU is usable with an \texttt{enhancement\_missing} indication.
It is not relabeled as a complete \ModeZero{} packet.

\section{Implant Transmitter Architecture}
\label{sec:implant-tx}

\subsection{Attribution and Architectural Boundary}

The implant endpoint connects the layered packet format in Section~\ref{sec:protocol} to a crystal-less IR-UWB pulse generator.
Its physical chain is
\begin{equation*}
\begin{split}
&\text{neural-data interface}\rightarrow\text{packet engine}
\rightarrow\text{lane encoders}\\
&\rightarrow\text{\ModeZero{}/\ModeOne{} mapper}\rightarrow
\text{RO/edge combiner}\rightarrow\text{pulse control}\\
&\rightarrow\text{PA}\rightarrow\text{matching/implant antenna}.
\end{split}
\end{equation*}
This block-level connection and its TX/RX protocol correspondence are part of
the present system architecture summarized in
Fig.~\ref{fig:end-to-end-architecture}.
The underlying free-running RO, selected-phase edge-combining pulse generator, and associated transmitter circuit topology are taken from Lei \emph{et al.} \cite{Lei2024}.
They are not claimed as original circuits here.
The discussion below is therefore an integration specification, not a transistor-level reproduction or a claim of independently measured transmitter performance.

\subsection{Packet and Lane Path}

The packet engine accepts framed neural data rather than raw electrode waveforms.
It serializes the robust Header and independently applies the declared AEAD, CRC, scrambling, FEC, and interleaving operations to Base and Enhancement PDUs.
The resulting coded streams are read in physical symbol order.
Bootstrap fields and the Header use the PWM Base path so that their interpretation does not depend on phase-lane acquisition.
Filler and pilots are generated deterministically at the packet layer.
Differential state is reset at packet initialization, \PhaseTrain{} entry,
payload entry, each pilot-island entry, and payload resumption after each
pilot island, as summarized with the receiver rule in Sec.~\ref{sec:receiver}.

Let $b_{w,k}$ denote the coded Base bit and $b_{\phi,k}$ the coded Enhancement bit for symbol $k$.
The width mapper selects a short or long integer-cycle burst according to $b_{w,k}$.
In \ModeZero{}, the polarity state is fixed and only $b_{w,k}$ conveys information.
In \ModeOne{}, the same width mapping remains present and the differential polarity state obeys
\begin{equation}
 p_k=p_{k-1}(-1)^{b_{\phi,k}},
 \qquad p_k\in\{-1,+1\},
 \label{eq:tx-dbpsk}
\end{equation}
after each explicit reset.
Thus \ModeOne{} adds a binary polarity choice; it does not require an implant I/Q DAC, quadrature mixer, or coherent absolute-phase reference.
Width and polarity controls are latched before the corresponding burst window and held constant during that window.

\subsection{Free-Running RO and Pulse Formation}

The adopted source architecture uses a nominal 300-MHz, 42-stage bi-phase RO
that exposes 84 phases; one of every three phases is selected so that 28 edge
events are available to the edge combiner in each transmitter cycle
\cite{Lei2024}.
The RO frequency is a nominal architectural parameter, not a measured value for a transmitter fabricated in this study.
Selected phase events define the burst boundaries, while edge combining produces the alternating high-rate drive presented to the pulse-control and PA path.
Changing the selected stop event changes the number of complete RF cycles and therefore implements PWM without changing the fixed symbol slot.
The differential polarity state selects the sign of the resulting burst for \ModeOne{}.
This manuscript does not specify transistor dimensions, physical phase routing, edge-combiner gate sizing, PA device segmentation, matching values, or antenna geometry.

The RO is modeled as free-running throughout a packet so that its timing and phase evolve continuously across preamble, training, Header, payload, and pilots.
It may be power-gated between packets, but symbol-by-symbol RO restart is excluded because it would break the assumed differential continuity.
After an interpacket restart, transmission begins only after an implementation-defined settling interval.
The PA is enabled only for the selected short or long burst window and remains off during the rest of the symbol slot.
RO supply isolation, startup time, PA switching transients, neural-front-end interference, peak-current delivery, spectral compliance, and thermal limits require circuit and measurement validation; none is inferred from the packet simulation.

\subsection{Normalized Spectral Sanity Check}

\begin{figure*}[t]
  \centering
  \includegraphics[width=0.96\textwidth]{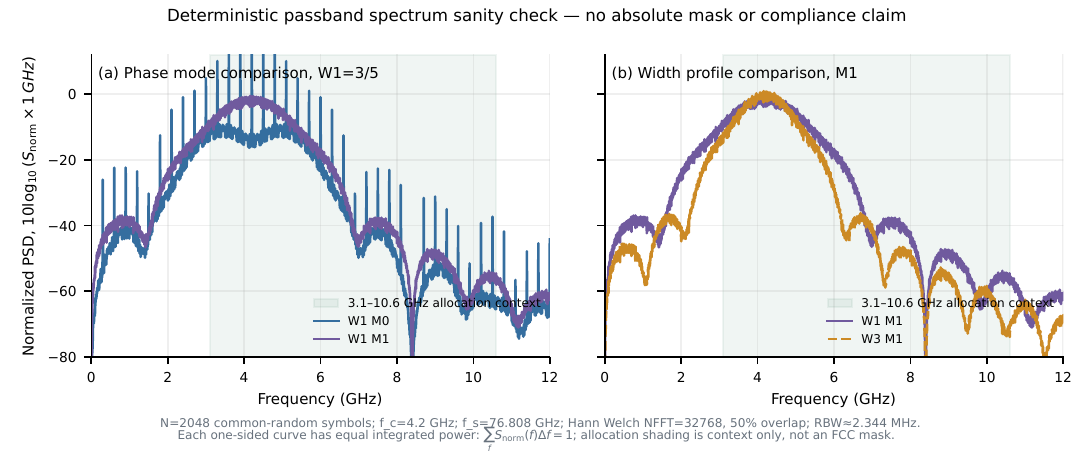}
  \caption{Normalized simulated TX power spectral density. The shaded
  3.1--10.6-GHz region is frequency-allocation context only. The curves are
  not in dBm/MHz and do not establish FCC-mask compliance, radiated power,
  antenna efficiency, or a measured implant output spectrum.}
  \label{fig:tx-normalized-psd}
\end{figure*}

Figure~\ref{fig:tx-normalized-psd} reports a deterministic long-stream
sample-domain spectral sanity check for the declared 300-MHz symbol rate and
4.2-GHz carrier.
Common randomized Base and Enhancement sequences isolate the effect of
\ModeZero{} versus \ModeOne{} differential polarity and W1 versus W3 pulse
width.
Hann-window Welch averaging is applied to the real passband synthesis, and
each curve is normalized to equal integrated power so that the comparison
addresses spectral shape rather than an uncalibrated PA amplitude.

\subsection{PA, Antenna, and Calibration Interface}

The pulse-control output drives a conceptual PA, matching network, package, and implant antenna whose aggregate effect enters the transmitted waveform and channel model.
The simulations do not claim a PA efficiency, radiated spectrum, output power, antenna gain, tissue range, or implant energy derived from new silicon.
Pulse width and polarity are protocol-controlled, whereas PA amplitude or power index is treated as configured metadata subject to implementation limits.

The technical model begins with an already authorized transmitter configuration and exposes packet-level RO variation to the receiver.
Wafer or post-package trim, reference-assisted TX calibration, calibration storage, production test, and any external command used to update a trim value are boundary conditions rather than proposed mechanisms.
In particular, the one-way architecture contains no implant receiver or external downlink, and receiver tracking cannot establish transmitter spectral compliance.

\section{Wearable External Receiver Architecture}
\label{sec:receiver}

\subsection{Signal-Path Partition}

The proposed wearable receiver separates always-required Base processing from optional Enhancement processing.
Its placement relative to the implant, transcutaneous channel, and host is
shown in Fig.~\ref{fig:end-to-end-architecture}; the present section expands
the wearable half of that system partition.
Its circuit-oriented dataflow is
\begin{equation*}
\begin{aligned}
\text{antenna}&\rightarrow\text{matching/BPF}\rightarrow\text{LNA/VGA AFE},\\
\text{AFE}&\rightarrow\text{envelope/integration Base path},\\
\text{AFE}&\rightarrow\text{gated burst-I/Q Enhancement path},\\
\text{both paths}&\rightarrow\text{digital baseband}\rightarrow\text{host}.
\end{aligned}
\end{equation*}
The notation identifies required functions and interfaces, not a completed transistor-level design.
After antenna filtering and gain control, the envelope path performs energy search, preamble correlation, timing observation, and short/long pulse discrimination.
A second, gated I/Q integrate-and-dump path yields one complex burst sample when phase processing is enabled.
Programmable integration windows and finite-resolution samples connect the AFE to the baseband; continuous multi-GS/s waveform export to the host is not assumed.
The architecture does not require a separate analog chain for each digital template.
Short/long envelope templates and the four adjacent-width transition classes used by differential phase processing are digital states applied to observations from the shared paths.

\begin{figure*}[t]
  \centering
  \includegraphics[width=0.95\textwidth]{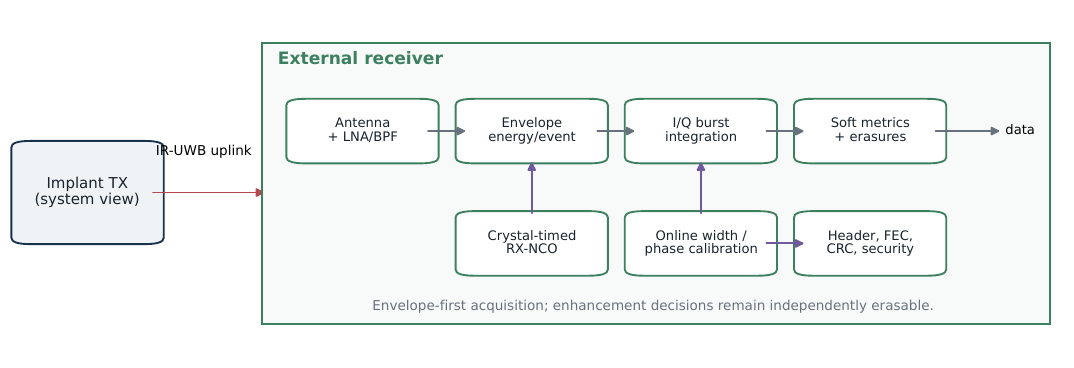}
  \caption{Wearable external-receiver partition. One envelope path supplies acquisition,
  timing, and Base-width evidence; one time-shared burst-I/Q path supplies
  optional differential-phase evidence. All templates and lane decisions are
  digital.}
  \label{fig:rx-architecture}
\end{figure*}

The wearable clock subsystem supplies the local RF/measurement reference, sampling clocks, and the digital-baseband clock.
Its precise local oscillator is a receive reference, not a mechanism that locks the implanted RO.
Acquisition estimates packet start and symbol period from multiple preamble events.
Thereafter, an NCO schedules each integration window:
\begin{align}
 \widehat t_{k+1}
 &=\widehat t_k+\widehat T_s[k+1]+K_p e_k,\\
 \widehat T_s[k+1]
 &=\widehat T_s[k]+K_i e_k ,
 \label{eq:nco}
\end{align}
where $e_k$ is admitted only when a bounded envelope correlation is reliable.
This is a causal arrangement: the predicted window collects the current burst, and the current timing observation corrects future windows.

\begin{figure}[t]
  \centering
  \includegraphics[width=\columnwidth]{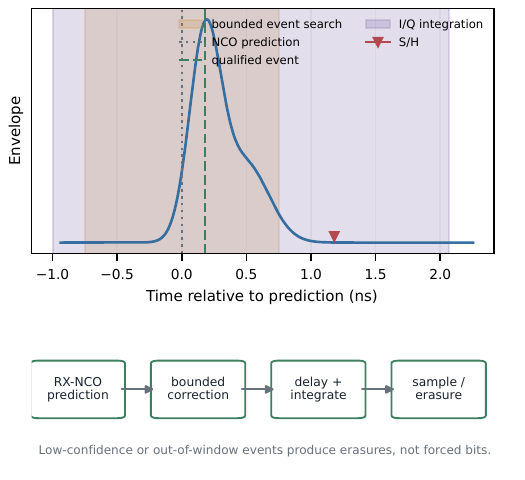}
  \caption{Predictive NCO sampling window. Envelope evidence from an admitted
  observation corrects future windows; it does not reveal the transmitted
  event time to the runtime detector.}
  \label{fig:rx-nco}
\end{figure}

\subsection{Envelope Evidence}

Let $\mathbf y_k\in\mathbb{R}^{64}$ be the time-resampled envelope-power row
for symbol $k$ after the causal AFE and AGC.
The implemented detector subtracts the row's 20th percentile and clips
negative samples:
\begin{equation}
 \widetilde{\mathbf y}_k
 =\max\{\mathbf y_k-Q_{0.2}(\mathbf y_k),0\}.
\end{equation}
Let $\boldsymbol\mu_0$ and $\boldsymbol\mu_1$ be the short- and long-burst
class means learned from valid \WidthTrain{} rows and define
\begin{align}
 d_b(k)&=\|\widetilde{\mathbf y}_k-\boldsymbol\mu_b\|_2^2,\\
 s_{\mathrm{shape},k}
 &=\operatorname{clip}_{24}\!
 \left[
 \frac{d_0(k)-d_1(k)}
 {\max\{2\,\operatorname{median}_{j\in\mathcal T}
          \min_b d_b(j),10^{-5}\}}
 \right],
 \label{eq:shape-score}
\end{align}
where $\mathcal T$ is the known width-training set and
$\operatorname{clip}_{a}(x)=\min(a,\max(-a,x))$.
For row energy $e_k=\mathbf 1^\mathsf T\widetilde{\mathbf y}_k$, the scalar
training statistic is
\begin{equation}
 s_{\mathrm{E},k}=\operatorname{clip}_{24}\!
 \left[
 \frac{(\bar e_1-\bar e_0)
       (e_k-\tfrac12(\bar e_0+\bar e_1))}
 {\max\{\tfrac12(v_0+v_1),10^{-5}\}}
 \right],
\end{equation}
where $\bar e_b$ and $v_b$ are the class sample mean and variance.
The online width soft value is
\begin{equation}
 L_{w,k}=\operatorname{clip}_{8}
 \left[2(0.75s_{\mathrm{E},k}+0.25s_{\mathrm{shape},k})\right].
 \label{eq:widthllr}
\end{equation}
There is no per-row amplitude fit, sample-weight mask, template shrinkage,
or separately calibrated residual variance in this frozen path.
Energy below threshold, clipping, an invalid window, or low $|L_{w,k}|$ produces an erasure or near-zero LLR rather than a forced bit.
The soft output is retained through deinterleaving and FEC decoding.

\subsection{Burst-I/Q Evidence}

For phase-enabled symbols, the gated I/Q path produces one complex observation $z_k$.
The mandatory baseline metric is a differential product,
\begin{equation}
 q_k=z_k z_{k-1}^{*},
\end{equation}
at symbol positions that do not reset the differential accumulator and for
which both adjacent observations are valid.
Because pulse width can alter the integrated complex response, the digital predictor is indexed by the adjacent Base-width class
$c_k\in\{00,01,10,11\}$.
The frozen receiver does not use transmitted-width truth and does not
marginalize over the four classes.
It makes the hard class selection
\begin{align}
 \widehat w_i&=\mathbf{1}\!\left[L_{w,i}\geq 0\right],\\
 \widehat c_k&=2\widehat w_{k-1}+\widehat w_k,
\end{align}
and applies the corresponding unit-magnitude phase-training template
$\mu_{\widehat c_k}$:
\begin{equation}
 L_{\phi,k}=
 -\kappa\,
 \frac{\Re\!\left\{q_k\mu_{\widehat c_k}^{*}\right\}}
 {|q_k|+\epsilon},
 \qquad \kappa=6.
 \label{eq:phase-hard-class}
\end{equation}
Here $\epsilon$ prevents division by zero.
When the boundary reset flag is set, either adjacent observation is invalid,
or $|L_{\phi,k}|$ is below the frozen erasure threshold, the output phase LLR
is set to zero.
If phase training, timing, or class evidence is inadequate, the Enhancement LLR is erased while envelope processing continues.

\begin{center}
\begin{minipage}{0.98\columnwidth}
\centering
\footnotesize
\textbf{Differential-state reset summary}\\[2pt]
\begin{tabular}{@{}p{0.30\columnwidth}p{0.62\columnwidth}@{}}
\toprule
Reset position & Transmitter/receiver action \\
\midrule
First ACQ symbol &
Initialize the packet differential accumulator; no cross-packet product. \\
First \PhaseTrain{} symbol &
Reset before the known training transition; assign no metric to this boundary symbol. \\
First payload symbol &
Apply the coded transition, but output a zero phase LLR because no cross-Header product is permitted. \\
First symbol of every eight-symbol pilot island &
Reset at pilot-island entry and assign no metric to the first known pilot transition. \\
First payload symbol after every pilot island &
Apply the coded transition, but output a zero phase LLR because no cross-pilot product is permitted. \\
\bottomrule
\end{tabular}
\end{minipage}
\end{center}

At a reset position, the transmitter clears the accumulator before applying
the current transition, while the receiver suppresses the differential metric
that would cross the boundary.
When that position belongs to the coded Enhancement payload, the symbol remains
in the transmitted lane but enters the soft FEC decoder with a zero LLR as a
deliberate boundary erasure; it is not relabeled as a pilot or reference
symbol.

\subsection{Digital Baseband and Host Boundary}

The digital baseband sequences acquisition, \WidthTrain{} and \PhaseTrain{} processing, SFD detection, Header decoding, lane demapping, deinterleaving, soft FEC decoding, descrambling, CRC checks, and authenticated decryption.
Header verification configures the remainder of the packet; a failed Header does not cause speculative profile or payload decoding.
Verified Base and Enhancement PDUs, sequence/session metadata, and staged receive status are delivered across the host interface.
For a valid Base PDU with missing Enhancement data, the interface emits an explicit \texttt{enhancement\_missing} indication.
It never promotes unverified samples or relabels the packet as transmitted \ModeZero{}.
Buffer sizing, bus protocol, clock-domain crossings, host processor choice, and software driver are implementation decisions outside this architectural study.

\subsection{Implementation Accounting}

The evaluation counts additions, multiplications, comparisons, stored coefficients, and peak soft-bit buffering.
These are architecture-level complexity measures.
Python execution time is not converted to ASIC power, and operation counts are not converted to gate area.
ADC resolution, template coefficient precision, and LLR precision will be swept to identify sensitivity without claiming a fabricated implementation.
No wearable-receiver sensitivity, noise figure, gain range, clock-jitter tolerance, power, area, or silicon timing result is claimed.

\section{Layered Protocol and Deterministic Processing}
\label{sec:protocol}

\subsection{Paired Modulation and Demodulation}

\begin{figure*}[t]
  \centering
  \includegraphics[width=0.96\textwidth]{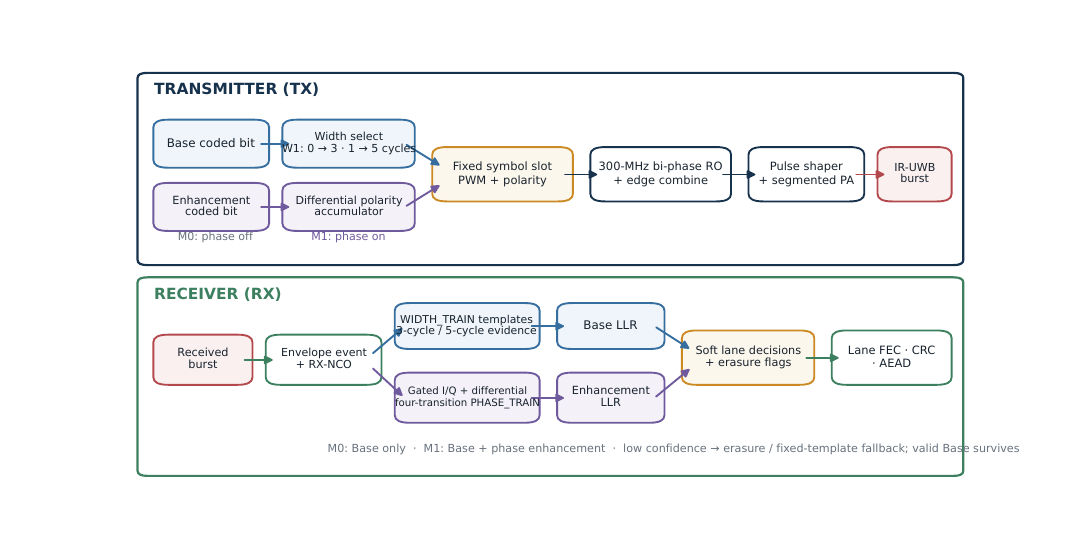}
  \caption{Logical transmit modulation and receive demodulation flow. This
  original block diagram follows the paired modulator/demodulator presentation
  convention of \cite{Lei2024}, while replacing its D16PPM hybrid modem with
  the present fixed-slot PWM Base lane (illustrated with the W1 3/5-cycle
  baseline), optional DBPSK Enhancement lane,
  predictive receive timing, soft LLRs, and explicit erasure/fallback path.}
  \label{fig:layered-modem-flow}
\end{figure*}

Figure~\ref{fig:layered-modem-flow} makes the physical correspondence explicit.
At the implant, the coded Base bit selects the short or long integer-cycle
burst, while the coded Enhancement bit updates the differential polarity state
only in \ModeOne{}.
At the wearable, the envelope/NCO/template branch produces the Base LLR and
the gated I/Q differential branch produces the optional Enhancement LLR.
These soft outputs are not application data: each lane must still pass the
declared decoder, CRC, and AEAD checks.

\subsection{Waveform-to-Soft-Metric Example}

\begin{figure*}[t]
  \centering
  \includegraphics[width=0.98\textwidth]{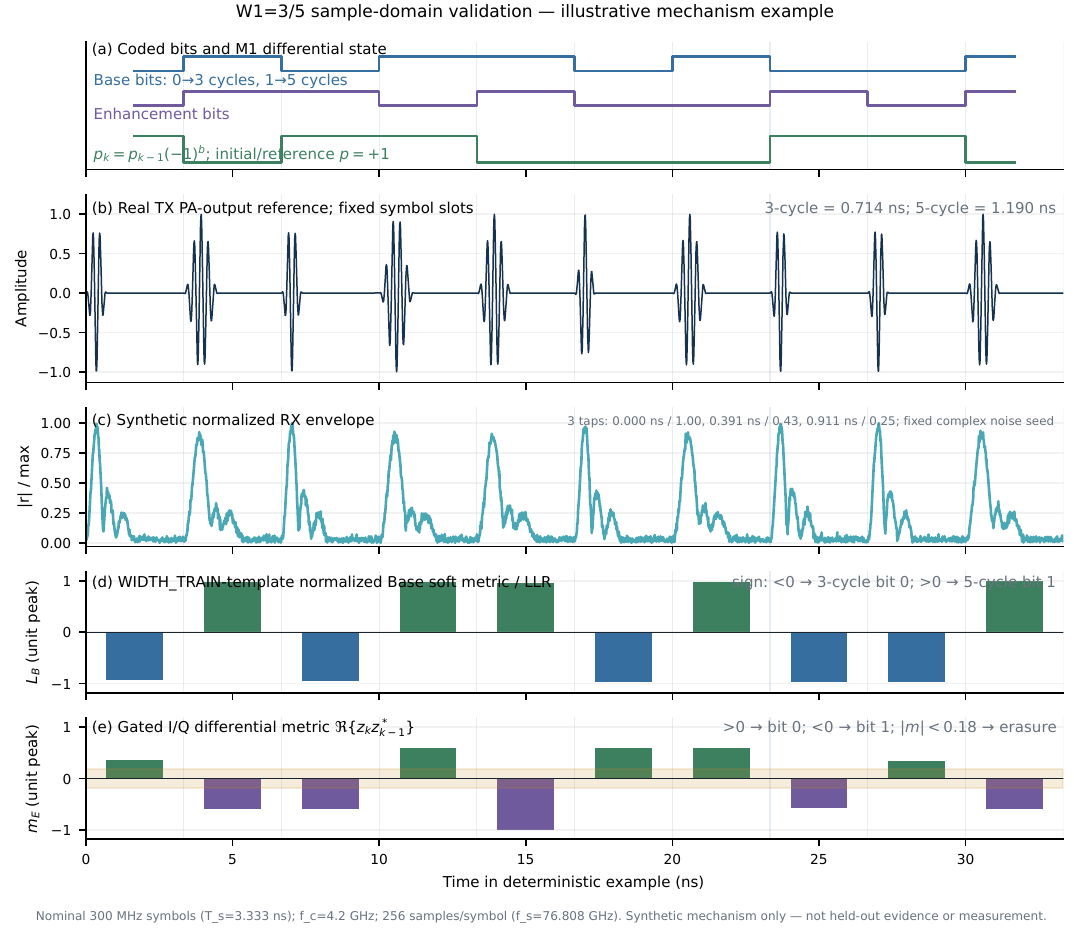}
  \caption{Deterministic W1/\ModeOne{} sample-domain modulation and
  demodulation example. Base bits select three- or five-cycle bursts;
  Enhancement bits update differential polarity; the synthetic receiver
  envelope and gated I/Q observations produce signed soft metrics. All
  amplitudes and metrics are normalized and no measured waveform is implied.}
  \label{fig:tx-rx-waveform}
\end{figure*}

The logical correspondence is complemented by the deterministic
sample-domain example in Fig.~\ref{fig:tx-rx-waveform}.
The figure applies W1 3/5-cycle PWM and the differential state update in
\eqref{eq:tx-dbpsk}, renders the real 4.2-GHz transmitter waveform inside
fixed 3.333-ns slots, and passes it through a declared synthetic demonstration
channel.
The receiver panels show the sampled envelope, a normalized short/long
template metric, and the gated I/Q differential product.
This example verifies sign, timing, pulse-width, and lane mapping; it is not a
packet-success result and is not drawn from a held-out test packet.

\subsection{Packet Order and Bootstrap}

The uplink packet is serialized in the following fixed order:
\begin{equation*}
\begin{split}
&\mathrm{ACQ}\rightarrow\WidthTrain\rightarrow\PhaseTrain\rightarrow
\mathrm{SFD}\\
&\quad\rightarrow\mathrm{ROBUST\_HEADER}\rightarrow
\mathrm{PAYLOAD/PILOTS}.
\end{split}
\end{equation*}
The provisional lengths are 64, 32, 32, and 32 symbols for ACQ, \WidthTrain{}, \PhaseTrain{}, and SFD, respectively.
All multibyte values use network byte order and fields are serialized most-significant bit first.
The width profile used to decode the bootstrap fields is known from authenticated session or scheduling context before Header decoding; the receiver does not try multiple profiles and accept whichever gives a valid CRC.
The implant packet engine emits this order without an out-of-band payload boundary, and the wearable state machine advances only after the corresponding acquisition, training, delimiter, or integrity condition succeeds.

\begin{figure*}[t]
  \centering
  \includegraphics[width=0.94\textwidth]{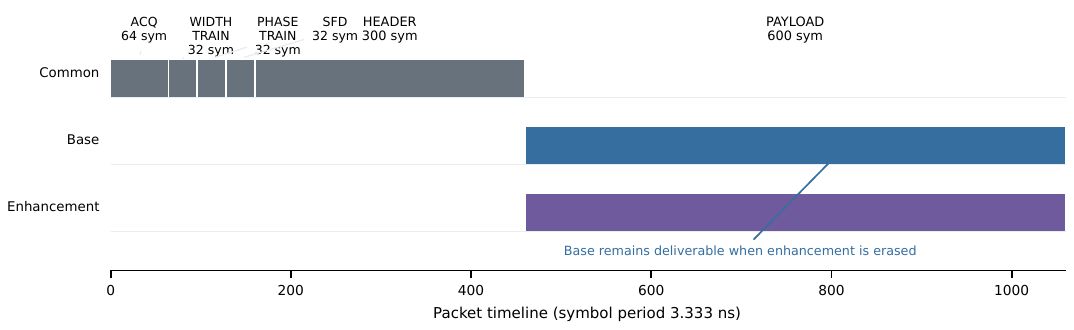}
  \caption{Packet timeline and the independently verified Base and Enhancement
  lanes. Enhancement failure can produce labeled Base-only service but cannot
  be counted as strict complete delivery.}
  \label{fig:packet-timeline}
\end{figure*}

\subsection{Robust Header}

The Header contains 128 information bits followed by CRC-16/CCITT-FALSE.
It is encoded by the rate-$1/2$, constraint-length-$7$ convolutional code with generators $(171_8,133_8)$ and six zero-tail input bits, yielding
\begin{equation}
 (128+16+6)\times 2=300
\end{equation}
PWM-only Header symbols.
Its bit fields are Version (4), Packet Type (4), PHY Mode (2), Base FEC (3), Enhancement FEC (3), Width Profile (4), TX Power Index (4), Pilot Interval Code (4), Flags (4), Base Length (16), Enhancement Length (16), Sequence Number (32), Short Session ID (16), Stream ID (8), and Key Epoch (8).
The CRC parameters are polynomial \texttt{0x1021}, initial value \texttt{0xFFFF}, no input/output reflection, and zero final XOR.

The Header is a hard interpretation gate.
Without a verified Header, the receiver cannot safely infer lane lengths, FEC, nonce context, or mapping.
Thus, even decodable payload samples are discarded after Header failure rather than being decoded under guessed metadata.
On transmit, the serialized information bits and CRC are convolutionally encoded, terminated, and mapped to PWM in the stated order.
On receive, PWM LLRs are passed to the soft convolutional decoder, the zero tail and field length are checked, and the reconstructed information is accepted only after CRC verification.

\subsection{Independent Lane Processing}

Base and Enhancement application PDUs are processed independently:
\begin{equation*}
\begin{split}
\mathrm{PDU}&\rightarrow\mathrm{AEAD}\rightarrow\mathrm{CRC32C}
\rightarrow\mathrm{scramble}\\
&\rightarrow\mathrm{FEC}\rightarrow\mathrm{interleave}
\rightarrow\mathrm{lane\ map}.
\end{split}
\end{equation*}
The inverse receive chain preserves soft information through FEC and performs CRC and authenticated-decryption checks before delivery.
The two lanes use separate scrambling, interleaving, FEC state, integrity result, and key-derivation domain.
Base coded bits map to pulse width; Enhancement coded bits map to DBPSK transitions.
If coded lengths differ, deterministic filler keeps the physical lanes aligned but is excluded from application integrity checks and delivery.

More explicitly, the wearable applies the inverse chain in physical receive order:
\begin{equation*}
\begin{split}
\mathrm{soft\ demap}&\rightarrow\mathrm{deinterleave}
\rightarrow\mathrm{FEC}^{-1}\rightarrow\mathrm{descramble}\\
&\rightarrow\mathrm{CRC32C\ verify}
\rightarrow\mathrm{AEAD\ verify/decrypt}.
\end{split}
\end{equation*}
A lane is deliverable only if its declared length, FEC termination or tail-biting condition, CRC, and AEAD result are all valid.
CRC supplies transmission-error detection within the simulation chain; AEAD supplies the protocol authentication boundary.
Neither check is replaced by a favorable demodulator metric.

\subsection{FEC Profiles and Soft Decoding}

The primary held-out campaign uses the same rate-$1/2$, constraint-length-$7$
convolutional mother code, $(171_8,133_8)$, for the Header and payload lanes.
The Header and baseline payload profiles append six zero-tail bits and use a
soft-input Viterbi decoder.
The sensitivity study additionally evaluates two protocol-selectable payload
profiles without changing the PWM/DBPSK mapper:
\begin{enumerate}
    \item a rate-$1/2$ tail-biting convolutional code (TBCC) using the same
    generator pair and a circular-state soft decoder, which removes tail
    overhead but does not add redundancy; and
    \item an effective rate-$1/4$ repeated-convolutional profile that transmits
    each mother-code output twice, sums the paired LLRs, and then applies the
    terminated soft Viterbi decoder.
\end{enumerate}
The repeated-convolutional Header uses the same LLR-combining principle.
Fixed two-copy robust profiles operate one level above FEC: the wearable sums
aligned LLRs from two complete scheduled packet copies before decoding, while
the implant always sends and is charged for both copies because the modeled
uplink has no acknowledgement path.
TBCC, repeated coding, and fixed-copy combining are therefore distinct
mechanisms and are reported with their actual airtime, goodput, energy, and
latency costs.

\subsection{Transmitter--Receiver Correspondence}

Every nonreset data-bearing transmit operation has a declared receive inverse;
the reset positions listed in Sec.~\ref{sec:receiver} are explicit zero-LLR
boundary erasures presented to the soft decoder.
The implant's fixed symbol order is consumed by the wearable packet state machine; its width selection is observed by the envelope branch; its differential polarity update in \eqref{eq:tx-dbpsk} is observed by the gated burst-I/Q differential product; and its field/pilot resets clear the receiver's corresponding differential history.
The known \WidthTrain{} width labels update only the two Base templates, while \PhaseTrain{} and pilot symbols provide known phase transitions for the Enhancement path.
Base-width uncertainty is retained as an LLR through deinterleaving and FEC.
For the separate phase metric, the signs of the two adjacent width LLRs select
one of four trained response classes according to
\eqref{eq:phase-hard-class}; transmitted-width truth is never used.
Header-declared mode and lengths then determine whether the receiver consumes a Base stream alone or aligned Base and Enhancement streams.

In \ModeZero{}, the implant emits no information-bearing Enhancement transitions and the wearable does not infer an Enhancement PDU.
In \ModeOne{}, complete delivery requires both independently encoded lanes to pass their receive checks.
The ability to deliver a valid Base PDU after Enhancement erasure is therefore a packet semantic designed into both endpoints, not a post hoc relabeling by the wearable.

The protocol exposes three noninterchangeable \ModeOne{} outcomes:
\begin{enumerate}
    \item complete: Header, Base, and Enhancement all verify;
    \item Base-only: Header and Base verify, Enhancement is absent or erased; and
    \item failure: the Header or Base does not verify.
\end{enumerate}
The second outcome carries an explicit missing-Enhancement indication.
It contributes to Base service statistics but not to \CompletePSR.
This distinction prevents a receiver-side degradation from being reported as a transmitted \ModeZero{} packet.

\subsection{Implementation Conformance Boundary}

The simulation uses paired encoder/decoder checks for Header serialization,
CRC-16, CRC-32C, convolutional termination, lane mapping, differential resets,
and pilot placement.
The manuscript specifies the processing order and transformations needed to
interpret the reported results, but it is not a complete interoperability
profile: a production specification must additionally freeze the AEAD suite
and nonce/AAD construction, key-derivation domains, scrambler polynomials and
seeds, interleaver permutations, filler generation, pilot sequences, and
published conformance vectors.
Production keys are necessarily outside the publication artifact.

\section{\WidthTrain{} Templates, Soft LLRs, and Gating}
\label{sec:training}

\subsection{Per-Packet Dual-Template Update}

The 32-symbol \WidthTrain{} field alternates known short and long widths, providing 16 labeled observations per class.
After NCO alignment, 64-point time resampling, validity screening, and
row-wise 20th-percentile subtraction, the external receiver computes
\begin{equation}
 \widehat{\boldsymbol\mu}_b
 =\frac{1}{|\mathcal V_b|}
 \sum_{k\in\mathcal V_b}\widetilde{\mathbf y}_k,
 \qquad b\in\{0,1\},
\end{equation}
where $\mathcal V_b$ contains valid training symbols of class $b$.
The frozen online receiver uses these class means directly:
$\boldsymbol\mu_b=\widehat{\boldsymbol\mu}_b$, equivalent to
$\lambda_0=\lambda_1=1$.
It does not fit a factory-shrinkage coefficient or a global residual
variance.
Instead, the packet-local median distance in \eqref{eq:shape-score}
normalizes the shape score, which is then fused with the pooled-variance
scalar-energy score according to \eqref{eq:widthllr}.
The known labels also permit a direct check of training re-decision error and LLR calibration.

The online templates target packet-specific amplitude, front-end response, and multipath shape.
They do not assert that physical PA tail or tissue propagation has been measured.
A fixed-template receiver remains a necessary baseline because it isolates the value of packet-local adaptation from the value of waveform correlation itself.
A scalar-width soft detector is a second baseline and can be safer than a mismatched detailed template.

\subsection{Training Quality Vector}

The gate evaluates a quality vector
\begin{equation}
 \mathbf q=
 [N_0,N_1,\rho_t,\rho_{\mathrm{clip}},D_{\mathrm{temp}},
 \mathrm{BER}_{\mathrm{train}}],
\end{equation}
where $N_b$ is the number of valid class-$b$ symbols, $\rho_t$ is timing-event validity, $\rho_{\mathrm{clip}}$ is clipping fraction, $D_{\mathrm{temp}}$ is normalized template separation, and $\mathrm{BER}_{\mathrm{train}}$ is known-symbol re-decision error.
Thresholds are selected on training/validation populations and frozen before test.
The final campaign used one immutable calibration artifact, identified in the
result manifest, for both held-out populations.
Numerically, acceptance requires $N_0,N_1\geq8$, $\rho_t\geq0.70$,
$\rho_{\mathrm{clip}}\leq0.20$, $D_{\mathrm{temp}}\geq0.80$, and
$\mathrm{BER}_{\mathrm{train}}\leq0.125$; these and the remaining receiver
constants are collected in Table~\ref{tab:frozen-receiver-parameters}.

Gating prevents an attractive but invalid failure mode: when a detailed template is wrong, decreasing noise can increase the magnitude of incorrect LLRs and contaminate the FEC decoder.
Therefore, training acceptance requires both sample sufficiency and evidence that the two learned classes remain separable without excessive clipping or timing loss.

\subsection{Profile-Aware Fallback}

The fallback is deterministic for a declared width profile:
\begin{equation*}
\begin{array}{ll}
\text{quality pass:} & \text{online template LLR},\\
\text{quality fail:} & \text{profile fallback, scaled LLR, erasure}.
\end{array}
\end{equation*}
The selected fallback and confidence scale are fixed from validation data.
They are not chosen using payload truth, test-set performance, or an undecoded Header.

For \ModeOne{}, failed width training also invalidates assumptions used by the phase path.
The receiver then suppresses Enhancement delivery and attempts only verified Base delivery using the conservative width fallback.
The same Base-only outcome is used if phase acquisition fails or if Enhancement FEC, CRC, or authentication fails after otherwise valid training.
At no point does the receiver manufacture missing Enhancement data or count the degraded packet as complete.

\begin{figure*}[t]
  \centering
  \includegraphics[width=0.84\textwidth]{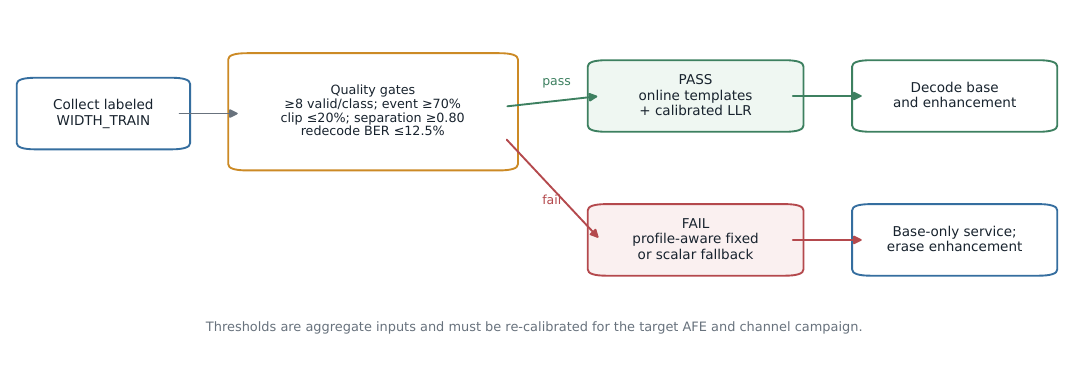}
  \caption{\WidthTrain{} quality gating and conservative fallback. The frozen
  campaign reports gated and ungated online-template receivers separately,
  because rejecting difficult packets must not be mistaken for improved
  strict delivery.}
  \label{fig:training-gate}
\end{figure*}

\section{Held-Out Evaluation Methodology}
\label{sec:methods}

\subsection{Publication Freeze}

The final campaign follows a train/validation/test split at the device/channel-cluster level.
Receiver thresholds, fusion weights, confidence scales, and fallback choices are selected without test access.
Before test execution, the campaign driver, waveform evaluator, configuration,
population manifests, and threshold file are hashed.
The result record stores these input hashes together with a root seed and a deterministic mapping from packet to device/channel cluster.
It does not contain a complete version-control snapshot of every imported
private module; this provenance limit is stated in
Sec.~\ref{sec:reproducibility}.

The waveform renderer uses RO event time, cumulative phase, fractional-delay multipath, causal filtering, gain control, ADC quantization, and clipping.
Capture, SFD, timing, envelope integration, and I/Q integration are derived from receiver samples.
True arrival time, true carrier offset, transmitted width labels outside training, and true payload boundaries are not available to the detector.
Sample-rate convergence of the event-timed complex-envelope renderer is
checked independently of the packet campaign.
The short modulation/demodulation trace and long-stream normalized PSD figures
are deterministic mechanism and spectral-sanity artifacts, respectively.
They use declared fixed sequences or seeds and are not included in PSR
estimators, confidence intervals, or held-out detector comparisons.
Their passband amplitudes are normalized; no simulated ordinate is converted
to dBm/MHz without a calibrated PA, load, package, and antenna reference plane.

\subsection{Paired Detector Comparison}

Every detector receives the same packet, RO trace, channel, noise realization, front-end samples, nominal energy, airtime, Header, and FEC.
Only the demapping algorithm changes.
The primary set comprises scalar-width soft detection, fixed dual templates, online dual templates, and quality-gated online templates.
A conventional noncoherent differential detector provides the phase-lane reference.

The main matrix covers \WidthOne{} and \WidthThree{}, \ModeZero{} and \ModeOne{}, nominal SNR conditions of 6, 9, 12, 15, and 18 dB, and rate-$1/2$ convolutional coding.
Additional coding and payload/pilot sweeps are secondary analyses.
Base-path renderer convergence is evaluated separately so that its numerical
fidelity and detector comparisons are not confounded.
\begin{table}[t]
\caption{Primary Held-Out Evaluation Matrix}
\label{tab:matrix}
\centering
\footnotesize
\begin{tabular}{@{}ll@{}}
\toprule
Dimension & Primary settings \\
\midrule
Width profile & \WidthOne{}, \WidthThree{} \\
PHY mode & \ModeZero{}, \ModeOne{} \\
Nominal SNR & 6, 9, 12, 15, 18 dB \\
Payload FEC & Conv.\ $K=7$, rate $1/2$ \\
Detector & Scalar, fixed, online, gated \\
Population & In-distribution and shifted hold-out \\
Packets/cell & 2000 (80,000 packet trials total) \\
Pairing & Common packet/front-end observations \\
\bottomrule
\end{tabular}
\end{table}

\subsection{Benign Reference Sensitivity and Robust Profiles}

A separate sensitivity campaign does not replace or retune the held-out
statistical tissue-proxy evaluation.
It retains the causal 256-sample/symbol receiver and the frozen online-template
calibration, but deliberately adopts a single-tap AWGN-like channel, no
shadowing or blocker, and 12-bit quantization.
This declared reference condition is closer to the assumptions used in
analytical or fixed ex-vivo link demonstrations, but it is not a human-channel
model.

Within this reference condition, paired common-random trials compare a normal
rate-$1/2$ convolutional Header/payload packet, a repeated-convolutional
Header, that Header with either tail-biting rate-$1/2$ convolutional payload
coding or repeated-convolutional effective rate-$1/4$ payload coding, and
fixed two-copy versions with soft LLR combining.
The two-copy modes have no implant feedback or early transmitter stop; both
back-to-back copies are always charged in energy, airtime, and delay.
The tabled TX-energy proxy is
\begin{equation}
 E_{\mathrm{TX,model}}=P_{\mathrm{ref}}\int |x_{\mathrm{norm}}(t)|^2\,dt,
\end{equation}
with the frozen configuration value $P_{\mathrm{ref}}=4.09$\,mW and the
normalized transmitted burst envelope $x_{\mathrm{norm}}(t)$.
It supports relative scheduling comparisons within this simulation only; it
is not a PA, antenna, battery, thermal, or measured implant-energy estimate.
Tail-biting coding removes termination overhead but is not described as
stronger redundancy than the terminated rate-$1/2$ code.

\subsection{Metrics and Uncertainty}

Primary packet metrics are Header PSR, Base PSR, Enhancement PSR, complete PSR, and Base-service PSR.
Pre-FEC width BER is reported only for packets reaching the corresponding observation stage and is not substituted for packet reliability.
For soft information, the study reports Brier score and a reliability curve; generalized mutual information is included when its estimator and conditioning are frozen.

Point estimates are accompanied by binomial intervals and cluster-resampled intervals.
Paired algorithm differences are resampled by device/channel cluster, and a paired binary test is used for packet-level discordance.
Zero observed failures are reported as zero observed failures with a one-sided upper confidence bound, never as true PER equal to zero.
The frozen campaign uses exactly \ResultPacketsPerCell{} packets in each of 40
cells: five SNR points, two width profiles, two modes, and two held-out
populations.
This sample size supports waterfall and algorithm-comparison estimates; it
does not certify sub-1\% PER.

\subsection{Complexity and Fidelity Checks}

For each receiver, the study records arithmetic operations per symbol, coefficient/state bytes, peak soft-buffer bytes, and burst-I/Q duty factor.
Renderer convergence compares 32, 64, 128, and 256 samples per symbol using
timing error, envelope-template distance, and pre-FEC Base-width BER.
The \ResultRendererRate{}-sample/symbol campaign setting is selected from
those Base/timing checks.
After the campaign freeze, a separate paired sensitivity check compares 256
and 512 samples/symbol for \ModeOne{} at 12 and 18\,dB on the synthetic
validation population.
It reuses each packet's payload, RO, channel, and noise seeds, applies the
256-sample frozen receiver calibration unchanged at both rates, and reports
Enhancement pre-FEC BER, phase-LLR agreement, and packet outcomes.
It is diagnostic only: it does not retune the receiver, consume either final
held-out population, or replace the 80{,}000-packet result.

\begin{figure}[t]
  \centering
  \includegraphics[width=\columnwidth]{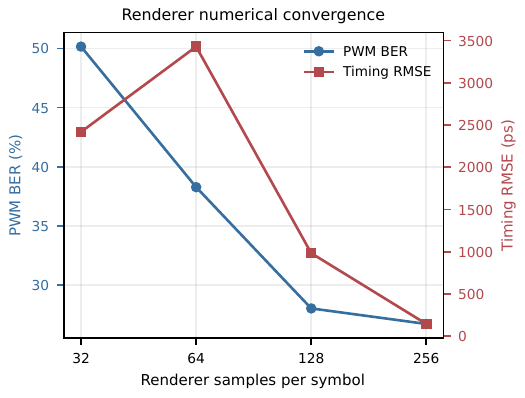}
  \caption{Renderer convergence. The selected 256-sample/symbol rate removes
  cycle slips in the convergence experiment and reduces sample-derived timing
  RMSE to 146\,ps; residual Base-width BER is therefore not attributed to
  renderer synchronization failure. DBPSK phase-metric convergence is not
  evaluated by this figure.}
  \label{fig:renderer-convergence}
\end{figure}

\begin{figure*}[t]
  \centering
  \includegraphics[width=0.88\textwidth]{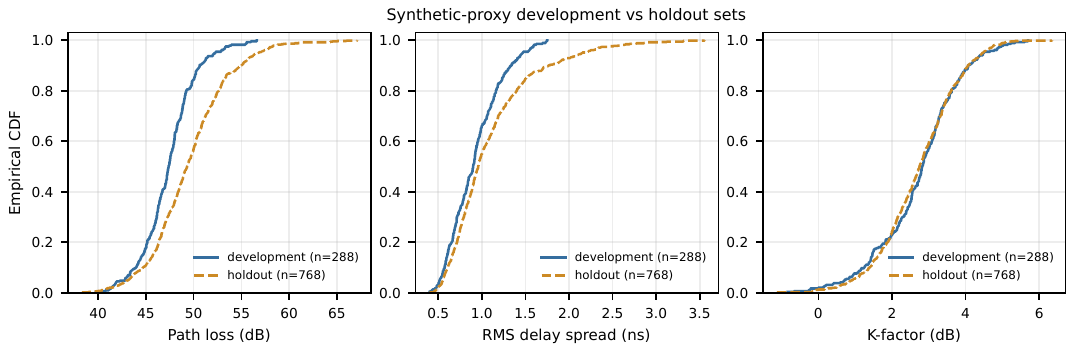}
  \caption{Synthetic tissue-proxy population distributions. These are
  generated statistical proxies, not human or animal measurements.}
  \label{fig:holdout-populations}
\end{figure*}

\section{Results and Complexity}
\label{sec:results}

The frozen record contains 80{,}000 packet trials with configuration hash
\ResultConfigHash{}.
Each of the 40 cells contains \ResultPacketsPerCell{} packets and 32
device/channel clusters.
All results below are synthetic simulation outcomes.

\subsection{Held-Out Waterfall}

\begin{figure*}[t]
  \centering
  \includegraphics[width=0.96\textwidth]{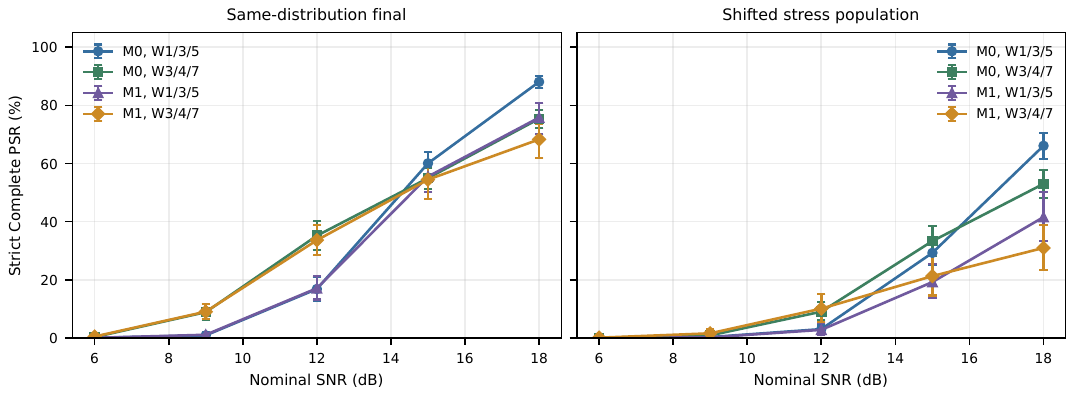}
  \caption{Strict Complete PSR of the ungated online-template receiver.
  Error bars are 95\% hierarchical cluster-bootstrap intervals. Nominal SNR
  is an integrated simulation reference, not TX power, RSSI, sensitivity, or
  range, and cannot be converted to those quantities without measured,
  jointly calibrated hardware and channel data.}
  \label{fig:waterfall}
\end{figure*}

Figure~\ref{fig:waterfall} shows a steep but incomplete waterfall.
In the same-distribution final population, M0/W1 rises from 0.85\% at
9\,dB to 60.1\% at 15\,dB and 88.1\% at 18\,dB.
M1/W1 reaches 75.8\% strict Complete PSR at 18\,dB; its Base service PSR is
81.5\%, so explicit degradation recovers 5.7 percentage points of verified
Base delivery without relabeling those packets as complete.
Neither result establishes product reliability.

The shifted loss/delay/blocker population is materially harder.
At 18\,dB, strict Complete PSR falls to 66.1\% for M0/W1 and 41.6\% for
M1/W1.
W3 is also consistently weaker than W1 at the highest tested SNR.
Table~\ref{tab:primary} gives the complete 18\,dB layer decomposition.

\begin{table*}[t]
\caption{Online-Template Held-Out Results at Nominal 18\,dB}
\label{tab:primary}
\centering
\footnotesize
\begin{tabular}{@{}lrrrrrrr@{}}
\toprule
Condition & Header & Base & Enh. & Strict Complete [95\% cluster CI] &
Service & Goodput & Width BER \\
\midrule
ID M0 W1 & 89.8 & 88.1 & -- & 88.1 [85.9, 90.2] & 88.1 & 42.6 & 4.46 \\
ID M0 W3 & 78.2 & 75.4 & -- & 75.4 [72.2, 78.5] & 75.4 & 36.4 & 5.99 \\
ID M1 W1 & 89.1 & 81.5 & 82.2 & 75.8 [70.0, 80.7] & 81.5 & 47.1 & 5.50 \\
ID M1 W3 & 79.0 & 70.2 & 75.6 & 68.3 [62.0, 73.7] & 70.2 & 42.0 & 7.08 \\
\midrule
Shift M0 W1 & 69.4 & 66.1 & -- & 66.1 [61.7, 70.5] & 66.1 & 31.9 & 11.01 \\
Shift M0 W3 & 56.8 & 53.0 & -- & 53.0 [48.2, 57.7] & 53.0 & 25.6 & 13.00 \\
Shift M1 W1 & 71.7 & 46.2 & 61.0 & 41.6 [33.4, 50.2] & 46.2 & 30.9 & 14.09 \\
Shift M1 W3 & 58.9 & 32.3 & 52.8 & 31.0 [23.3, 38.8] & 32.3 & 24.5 & 17.07 \\
\bottomrule
\end{tabular}
\vspace{2pt}

\raggedright\scriptsize
Percentages use 2000 packets per condition. ``ID'' is the frozen
same-distribution synthetic population; ``Shift'' is the predeclared
loss/delay/blocker stress population. Goodput is Mbit/s. Service includes
verified Base-only delivery; Strict Complete does not.
Nominal SNR is the internal $E_{\mathrm{ref}}/N_0$ renderer control defined
in \eqref{eq:snr-reference}; it is not $E_b/N_0$, RSSI, sensitivity, TX power,
or range and is not externally convertible without calibrated hardware and
channel measurements.
\end{table*}

\subsection{What Online Training and Gating Actually Changed}

Across the complete predeclared sweep, the online dual-template receiver has
a 24.4\% macro-average strict Complete PSR, compared with 21.7\% for scalar
width and 16.2\% for the frozen detailed template.
At 18\,dB in the same-distribution population, its four-cell macro-average
strict Complete PSR is 76.9\%, versus 73.2\% for scalar and 63.1\% for the
fixed template.
Thus packet-local adaptation is useful, but it does not remove the
classification/channel floor.

The quality gate accepts 70.9\% of packets over the full sweep.
Its fallback reduces macro-average strict Complete PSR from 24.4\% to
21.6\%; at 18\,dB in the same-distribution population it reduces the
four-cell average from 76.9\% to 73.3\%.
The tested gate is therefore a conservative failure-semantics mechanism, not
a reliability improvement.
It remains useful for preventing overconfident delivery, but should not be
advertised as PSR gain.

\begin{figure*}[t]
  \centering
  \includegraphics[width=0.94\textwidth]{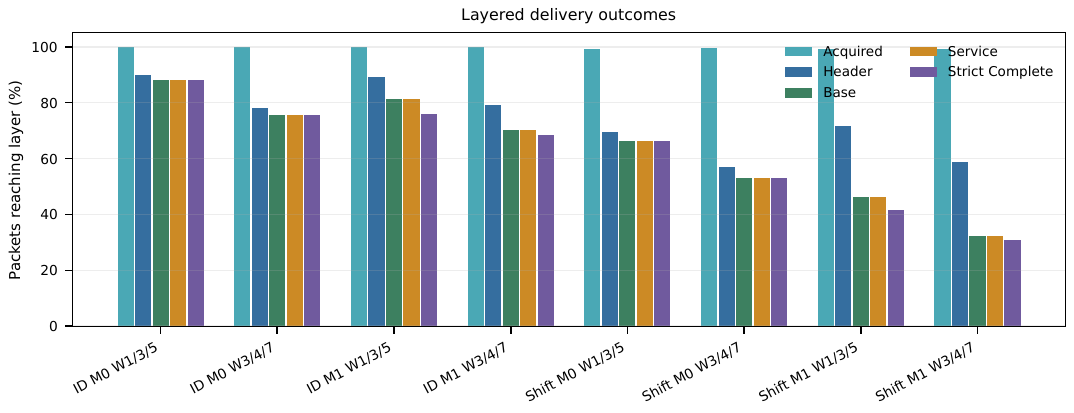}
  \caption{Layer reachability for the online-template receiver at nominal
  18\,dB. Service can exceed strict Complete only when a verified Base lane is
  delivered with Enhancement explicitly missing. The nominal SNR is the
  internal simulation reference of \eqref{eq:snr-reference}, not a real-world
  link-budget point.}
  \label{fig:layered-outcomes}
\end{figure*}

\subsection{Soft Information and Architecture Cost}

The sweep-aggregate Brier score improves from 0.173 for scalar width to 0.158
for online templates.
The corresponding unoptimized mismatched-LLR information-score proxy rises
from 0.132 to 0.250.
The fixed template has a negative aggregate score ($-1.36$), directly
exposing confident model mismatch rather than hiding it behind hard BER.

\begin{figure*}[t]
  \centering
  \includegraphics[width=0.82\textwidth]{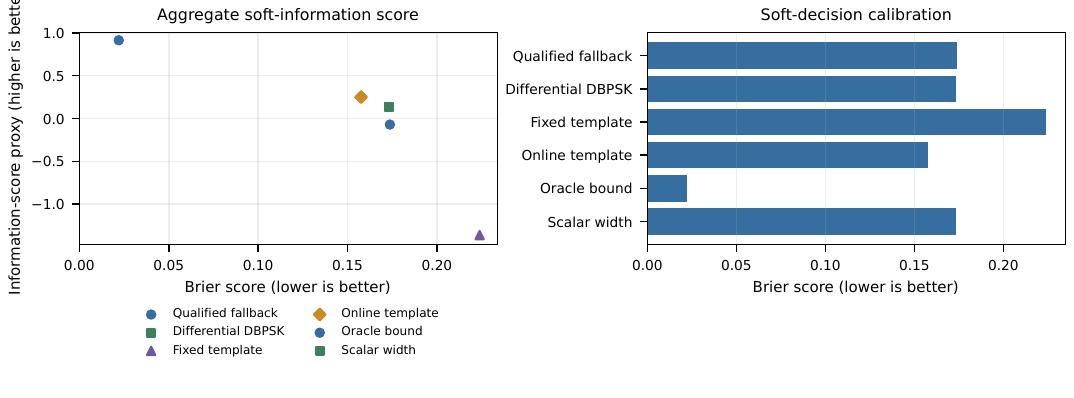}
  \caption{Aggregate soft-information evidence. Negative mismatched-LLR
  information scores indicate harmful confidence scaling; they are retained
  rather than clipped from the evidence.}
  \label{fig:llr-quality}
\end{figure*}

\begin{table}[t]
\caption{Deployable Detector Complexity and Sweep-Aggregate Outcome}
\label{tab:complexity}
\centering
\footnotesize
\begin{tabular}{@{}lrrrr@{}}
\toprule
Receiver & Ops/sym. & State B & PSR & Brier \\
\midrule
Scalar-width soft LLR & 18 & 32 & 21.7 & 0.173 \\
Fixed dual template & 280 & 512 & 16.2 & 0.224 \\
Online dual template & 320 & 1024 & 24.4 & 0.158 \\
Quality-gated online & 360 & 1152 & 21.6 & 0.174 \\
\bottomrule
\end{tabular}
\vspace{2pt}

\raggedright\scriptsize
PSR is the macro-average strict Complete outcome across both populations, all
five SNR points, two modes, and two width profiles; it is not a reliability
claim for any one operating point. Counts cover arithmetic and stored
coefficients, not gates, cycles, area, or power.
\end{table}

The online method uses \ResultOpsPerSymbol{} estimated operations per symbol
and \ResultStateBytes{} bytes of coefficient/state storage.
These values describe algorithmic accounting only.
They do not establish silicon area, clock rate, power, or energy.

\subsection{Benign Reference Robust Sensitivity}

Table~\ref{tab:robust-reference-results} gives the separate single-tap,
12-bit reference sensitivity campaign.
It uses 500 packets per cell and the same frozen online-template calibration;
it does not replace the held-out tissue-proxy results above.
At 12\,dB, no tested single-copy profile exceeds 9.2\% Complete PSR in M0 or
7.8\% in M1.
Fixed two-copy soft combining with a repeated Header and TBCC payload raises
the point estimates to 85.0\% and 91.2\%, respectively.
Repeating the payload code as well raises them to 92.2\% and 93.2\%, but
reduces goodput to 11.90 and 14.59\,Mbit/s and increases modeled transmit
energy and latency beyond the TBCC two-copy profile.

At 15\,dB, the normal single-copy profile reaches 85.4\% in M0 and 91.8\% in
M1, while both two-copy profiles observe 500/500 successes.
At 18\,dB, the normal profile also observes 500/500 and has the highest
goodput, 48.32\,Mbit/s in M0 and 57.60\,Mbit/s in M1.
Thus fixed repetition is useful in the waterfall region but is dominated at
high SNR.
The 500/500 observations have a 95\% Wilson lower bound of approximately
99.24\%; they are not evidence of zero true PER.
TBCC is treated as a rate-$1/2$ termination-efficiency option, not as
stronger redundancy than the terminated rate-$1/2$ convolutional code.

\begin{table*}[t]
\centering
\caption{Robust reference sensitivity results at selected nominal SNR values.}
\label{tab:robust-reference-results}
\footnotesize
\begin{tabular}{@{}cc l r r r r@{}}
\toprule
Mode & SNR (dB) & Variant & Complete PSR (\%) & Goodput (Mbit/s) & Model energy (nJ) & Latency ($\mu$s) \\
\midrule
M0 & 12 & Baseline conv 1/2 & 0.0 & 0.00 & 2.088 & 3.974 \\
M0 & 12 & TBCC + fixed blind $\times$2 & 85.0 & 16.54 & 5.146 & 9.867 \\
M0 & 12 & Repeated conv + fixed blind $\times$2 & 92.2 & 11.90 & 7.880 & 14.881 \\
\addlinespace
M0 & 15 & Baseline conv 1/2 & 85.4 & 41.27 & 2.088 & 3.973 \\
M0 & 15 & TBCC + fixed blind $\times$2 & 100.0 & 19.46 & 5.145 & 9.866 \\
M0 & 15 & Repeated conv + fixed blind $\times$2 & 100.0 & 12.90 & 7.879 & 14.879 \\
\addlinespace
M0 & 18 & Baseline conv 1/2 & 100.0 & 48.32 & 2.087 & 3.973 \\
M0 & 18 & TBCC + fixed blind $\times$2 & 100.0 & 19.46 & 5.145 & 9.867 \\
M0 & 18 & Repeated conv + fixed blind $\times$2 & 100.0 & 12.90 & 7.879 & 14.880 \\
\addlinespace
\midrule
M1 & 12 & Baseline conv 1/2 & 0.0 & 0.00 & 1.742 & 3.333 \\
M1 & 12 & TBCC + fixed blind $\times$2 & 91.2 & 20.52 & 4.443 & 8.533 \\
M1 & 12 & Repeated conv + fixed blind $\times$2 & 93.2 & 14.59 & 6.468 & 12.266 \\
\addlinespace
M1 & 15 & Baseline conv 1/2 & 91.8 & 52.88 & 1.742 & 3.333 \\
M1 & 15 & TBCC + fixed blind $\times$2 & 100.0 & 22.50 & 4.443 & 8.533 \\
M1 & 15 & Repeated conv + fixed blind $\times$2 & 100.0 & 15.65 & 6.466 & 12.266 \\
\addlinespace
M1 & 18 & Baseline conv 1/2 & 100.0 & 57.60 & 1.742 & 3.333 \\
M1 & 18 & TBCC + fixed blind $\times$2 & 100.0 & 22.50 & 4.439 & 8.533 \\
M1 & 18 & Repeated conv + fixed blind $\times$2 & 100.0 & 15.65 & 6.461 & 12.266 \\
\bottomrule
\end{tabular}
\begin{minipage}{0.98\textwidth}
\footnotesize Complete PSR is the point estimate; goodput and resource values charge all fixed blind copies. Model TX energy uses the normalized-burst proxy defined in Sec.~\ref{sec:methods}; it is not measured implant energy. Nominal SNR is the internal $E_{\mathrm{ref}}/N_0$ control of \eqref{eq:snr-reference}, not $E_b/N_0$, RSSI, sensitivity, TX power, or range; conversion requires calibrated hardware and channel measurements. Results are from the declared synthetic single-tap sensitivity model, not measured hardware or clinical evidence.
\end{minipage}
\end{table*}

\subsection{Paired DBPSK Sampling Sensitivity}

Table~\ref{tab:dbpsk-sampling-sensitivity} reports the separate 256/512
samples/symbol check.
Each cell contains 64 paired \ModeOne{} packets from the synthetic validation
population; neither final population is reused.
At 18\,dB, W3 is the most stable tested condition: Enhancement pre-FEC BER is
1.86\% at 256 and 1.88\% at 512 samples/symbol, the non-erased phase-LLR
correlation is 0.972, and hard-decision disagreement is 1.96\%.
W1 at the same SNR changes from 2.81\% to 3.65\% BER, with correlation 0.942
and 3.88\% disagreement.

At 12\,dB the corresponding correlations fall to 0.793 for W1 and 0.867 for
W3, and sign disagreement rises to 12.88\% and 8.90\%.
Enhancement-PSR point estimates also move in both directions across rates,
but 64 packets/cell are insufficient for a reliability comparison.
Thus the experiment detects material sampling sensitivity near the waterfall
and does not establish convergence of the frozen 256-sample phase path.
The pre-FEC labels are used only for post-decision scoring; they do not enter
template selection or demodulation.

\begin{table}[t]
\centering
\caption{Paired DBPSK sampling-rate sensitivity.}
\label{tab:dbpsk-sampling-sensitivity}
\scriptsize
\begin{tabular}{@{}ccrrrr@{}}
\toprule
Width & SNR & \multicolumn{2}{c}{Enh. BER (\%)} & LLR $\rho$ & Disagree. (\%) \\
\cmidrule(lr){3-4}
 & (dB) & 256 & 512 & 256/512 & 256/512 \\
\midrule
W1 & 12 & 9.52 & 9.15 & 0.793 & 12.88 \\
W1 & 18 & 2.81 & 3.65 & 0.942 & 3.88 \\
W3 & 12 & 8.19 & 7.33 & 0.867 & 8.90 \\
W3 & 18 & 1.86 & 1.88 & 0.972 & 1.96 \\
\bottomrule
\end{tabular}
\begin{minipage}{0.98\columnwidth}
\scriptsize Pre-FEC coded-Enhancement metrics; 64 paired validation packets/cell. The same payload, RO, channel, and noise seeds and frozen 256-sample calibration are used at both rates. $\rho$ and disagreement condition on both LLRs being non-erased. A shared noise seed does not create a sample-identical continuous-time noise realization on different grids.
\end{minipage}
\end{table}

\subsection{Interpretation}

The remaining acquisition, Header, and payload losses show that deployable
waveform classification and its interaction with channel/timing uncertainty
remain the principal receiver-design targets.
The results support the receiver partition, deterministic processing chain,
and explicit degradation semantics as research artifacts, but they do not
validate a commercial telemetry link.
No nominal SNR point is translated into TX power, receiver sensitivity,
tissue range, or a medical reliability target without a measured reference
plane and link budget.

\section{Limitations and Publication Boundary}
\label{sec:limitations}

This work is a pre-silicon simulation study.
It reports no fabricated transmitter or receiver, no measured receiver sensitivity, no measured power or area, and no post-layout result.
Its statistical channel is a tissue-motivated proxy rather than a measured human or animal channel population.
There is no human-subject or animal experiment, no clinical endpoint, and no validation of safety, efficacy, biocompatibility, SAR, temperature rise, electromagnetic compatibility, coexistence, cybersecurity compliance, or regulatory classification.
Accordingly, the results support architecture, receiver, and protocol design decisions only within the stated simulation assumptions.

The current front-end evidence must still be replaced or calibrated with measured antenna, package, channel, PA-burst, filter, gain-control, blocker, quantizer, and clipping data.
The deterministic TX waveform and equal-integrated-power PSD plots validate
mapping and spectral-shape calculations only.
They do not establish absolute conducted or radiated dBm/MHz, FCC-mask
compliance, PA efficiency, antenna efficiency, coexistence, or tissue
exposure.
The evaluation describes architectural operation counts and state, but these do not establish clock rate, memory macro choice, PVT margin, physical area, or energy.
Likewise, a successful CRC/authentication check in Monte Carlo does not estimate the cryptographic undetected-error probability or complete a security assessment.
The 2000-packet cell size estimates the observed waterfall but does not
certify sub-1\% PER.
ADC and fixed-point word-length sensitivity remain implementation-stage
validation tasks; no word-length freeze is claimed.
The renderer-convergence study validates Base-envelope timing and width
metrics only.
The added 256/512 paired DBPSK check narrows that gap but does not prove
continuous-time convergence.
The two grids use the same pseudorandom noise seed rather than a nested
continuous-time noise realization, per-sample AGC attack/release coefficients
remain unchanged, the 512-point path retains calibration fitted at 256
samples/symbol, and no 1024-point reference is included.
It therefore measures sensitivity of the complete discrete-time receiver
under one small synthetic validation design.
The held-out \ModeOne{} Enhancement and Complete PSRs remain conditional on
the frozen 256-sample implementation.
Under the unchanged channel model, memoryless envelope-template detection
retains substantial packet losses; future work must test measured guard-time
requirements or richer complex-sample channel training and equalization rather
than extrapolating the present curve.
The separate single-tap, 12-bit benign reference campaign is a sensitivity
bound for coding and fixed-copy policy comparisons.  Its higher packet success
must not be substituted for the held-out statistical tissue-proxy result,
interpreted as a measured ex-vivo reproduction, or used to claim product
reliability.

The transmitter attribution is intentionally narrow.
The free-running 42-stage bi-phase RO, selected-phase edge combining, and implant pulse-generation circuit are taken from Lei \emph{et al.} \cite{Lei2024}.
The present contribution is the end-to-end architectural partition that connects this attributed transmitter to the packet/lane mapper, channel model, wearable receiver, packet semantics, training adaptation, fallback, and evaluation framework.
It is not a claim of a new RO, edge combiner, PA, or measured transmitter implementation.
A reference-assisted transmitter-calibration option is compatible with the interface, but its algorithm, circuits, command transport, and implementation are outside this paper.
The manuscript does not reproduce unpublished circuit schematics, calibration implementation, device dimensions, internal register maps, production limits, or patent conclusions.

The reproducibility disclosures and the limits imposed by private code and
raw data are stated explicitly in Sec.~\ref{sec:reproducibility}.

An implant receiver, external downlink transmitter, and reference-assisted calibration link remain outside the technical design presented here.
Future work may consider those functions only after their architecture, safety, authentication, and power boundaries are defined independently.

\section{Reproducibility and Code Availability}
\label{sec:reproducibility}

\subsection{Frozen Parameters and Statistical Model}

Table~\ref{tab:frozen-receiver-parameters} records the effective values used
by the formal 80{,}000-packet campaign rather than defaults that were present
but inactive in other receiver prototypes.
In particular, the online templates are unshrunk class means, the shape
statistic uses a packet-local median-distance normalization rather than a
global fitted residual variance, and the quality-gate training-BER limit is
0.125.
\begin{table*}[t]
\centering
\caption{Frozen campaign and receiver parameters used for the reported held-out results.}
\label{tab:frozen-receiver-parameters}
\scriptsize
\begin{tabular}{@{}p{0.20\textwidth}p{0.76\textwidth}@{}}
\toprule
Parameter group & Frozen value or rule \\
\midrule
Campaign &
Hash \texttt{593f2287c9b49ead}; root seed 2026082699;
40 cells; 2000 packets/cell; 4000 cluster-bootstrap replicates;
sequential stopping disabled. \\
Calibration &
Hash \texttt{48cdcc968e591308}; 250 training and 250 validation
packets per algorithm; one immutable artifact applied to both final
populations. \\
Renderer and AFE &
300-MHz symbol rate; 256 samples/symbol (76.8-Gsample/s numerical
grid); 6-bit uniform I/Q quantization over $[-1,1]$; causal one-pole
750-MHz front end; AGC target 0.35, gain range 0.25--512, frozen after
64 symbols. \\
Acquisition and timing &
Acquisition correlation $\geq0.10$; refined SFD correlation $\geq0.30$;
event peak/noise $\geq3$; search half-window 0.75\,ns;
$K_p=0.10$, $K_i=0.002$; timing-error clamp $\pm0.10T$;
integration interval $[-0.30T,+0.62T]$ about the prediction. \\
Online width rows &
64 resampled envelope-power values; subtract row 20th percentile and
clip below zero; direct valid-class means
($\lambda_0=\lambda_1=1$); no per-row amplitude fit or sample weights. \\
Width statistic &
Shape-distance normalization by
$\max\{2\,\operatorname{median}_{\mathcal T}\min_b d_b,10^{-5}\}$;
pooled scalar-variance floor $10^{-5}$; component clipping $\pm24$;
fusion $0.75s_{\mathrm E}+0.25s_{\mathrm{shape}}$; final scale 2 and
clipping $\pm8$.  No separately calibrated $\sigma^2$ is used. \\
Training-quality gate &
$N_0,N_1\geq8$; timing-event rate $\geq0.70$; clipping fraction
$\leq0.20$; normalized template separation $\geq0.80$; known-symbol
re-decision BER $\leq0.125$. \\
Fallback &
Gate pass: online statistic.  Gate fail: $0.55$ times fixed-template
statistic for W1, or $0.55$ times scalar statistic for W3; final scale
2 and clipping $\pm8$. \\
Header confidence threshold &
Scalar 4.8532; fixed 7.4957; online 4.4043; adaptive 5.3264;
conventional differential 4.8532. \\
Base/phase erasure threshold &
Scalar 0.7156/0.7156; fixed 0.6500/0.0272; online
0.7131/0.7131; adaptive 0.6500/0.2389; conventional differential
0.7156/0.7156.  The Base value is
$\max(0.65,\text{calibrated threshold})$; the phase value is the
calibrated threshold directly. \\
DBPSK phase metric &
Four unit-magnitude adjacent-width templates; missing-class fallback to
the global phase template; phase-training coherence $\geq0.18$;
hard width-class selection; $\kappa=6$; boundary/invalid/low-magnitude
metrics erased. \\
\bottomrule
\end{tabular}
\end{table*}

Table~\ref{tab:frozen-stochastic-model} specifies the distributions, draw
levels, and within-packet correlations of the RO and channel perturbations.
All Gaussian variables listed there are mutually independent unless a shared
sampling level or accumulated process is stated.
The populations are synthetic design proxies, not fitted human or animal
measurements.
No result should be interpreted as estimating a biological population
distribution.
\begin{table*}[t]
\centering
\caption{Frozen stochastic RO and tissue-proxy model.}
\label{tab:frozen-stochastic-model}
\scriptsize
\begin{tabular}{@{}p{0.20\textwidth}p{0.48\textwidth}p{0.26\textwidth}@{}}
\toprule
Term & Distribution or deterministic rule & Sampling/correlation level \\
\midrule
RO residual &
$A_{\mathrm{dev}}\sim\mathcal N(0,6000^2)$ ppm &
One draw per packet; shared by every symbol. \\
Startup and drift &
$A_{\mathrm{start}}\sim\mathcal N(0,2500^2)$ ppm with
$e^{-(k+96)/48}$ decay;
$A_{\mathrm{end}}\sim\mathcal N(0,220^2)$ ppm with linear packet ramp &
Independent packet draws; correlated over symbols through the stated curves. \\
RO random walk &
$\eta_k\overset{\mathrm{iid}}{\sim}\mathcal N(0,2.5^2)$
ppm/symbol and $\epsilon_{\mathrm{rw}}[k]=\sum_{i=0}^k\eta_i$ &
Independent increments; accumulated within a packet. \\
Data pushing &
$120(w_k-\bar w)$ ppm &
Deterministic from the packet width sequence. \\
Timing/width perturbations &
Boundary jitter $\mathcal N(0,4^2)$ ps; packet-shared width offset
$\mathcal N(0,8^2)$ ps; per-symbol width jitter
$\mathcal N(0,2^2)$ ps; RX-LO offset $\mathcal N(0,2^2)$ ppm &
Independent draws except for the packet-shared width term; the RO state jointly
sets symbol period, carrier offset, width, and accumulated phase. \\
\midrule
ID proxy hierarchy &
Cluster path loss $\mathcal N(48,3^2)$ dB; proxy increment
$\mathcal N(0,1.5^2)$ dB; packet increment $\mathcal N(0,0.6^2)$ dB.
Cluster RMS delay $0.85e^{\mathcal N(0,0.30^2)}$ ns and proxy multiplier
$e^{\mathcal N(0,0.12^2)}$. &
32 final clusters, eight proxies/cluster; cluster and proxy terms recur,
packet terms are redrawn. \\
Shifted proxy hierarchy &
As above with cluster path loss $\mathcal N(53,4.5^2)$ dB, proxy
increment $\mathcal N(0,2.0^2)$ dB, and cluster RMS delay
$1.45e^{\mathcal N(0,0.42^2)}$ ns. &
32 final clusters, eight proxies/cluster; predeclared loss/delay shift. \\
Blocker and diversity &
ID: blocker Bernoulli 0.03 with 6-dB penalty and
$\max\{0,\mathcal N(2.2,0.8^2)\}$-dB diversity.
Shifted: 0.10, 9\,dB, and
$\max\{0,\mathcal N(1.2,1.1^2)\}$ dB. &
Redrawn per packet. \\
Four-tap channel &
Delays span 0 to $2.5$ times proxy RMS delay with nonzero-tap
$\mathcal N(0,0.06^2)$-ns jitter; decay
$\operatorname{clip}(\mathcal N(4.5,0.8^2),1.5,8)$ dB/tap;
tap-power jitter $\mathcal N(0,0.5^2)$ dB; independent uniform phases,
then unit-energy normalization. &
Decay is proxy-shared; delay/power perturbations and phases are packet-level. \\
\bottomrule
\end{tabular}
\end{table*}

\subsection{Code and Data Availability}

The simulation implementation, detailed configuration files, and packet-level
raw traces are not publicly released at submission because they remain under
intellectual-property, institutional, and collaborator review.
No release date is promised.
To make the evidentiary boundary explicit despite that restriction, this
paper publishes the effective campaign dimensions, receiver constants,
statistical distributions, reset and decision rules, result tables, and
cryptographic hashes of the campaign and frozen calibration artifacts.

These disclosures support audit of the stated model and internal consistency,
but they are not equivalent to executable artifact availability and do not
permit exact independent rerunning of the Monte Carlo campaign.
The archived result manifest hashes selected drivers, evaluator files,
configurations, population manifests, and calibration input, but not a
complete version-control snapshot of every imported private module.
The submission package therefore contains the manuscript source,
bibliography, vector figures, and static result tables only; it excludes
private code, configurations, raw packet records, credentials, and internal
design material.

\section{Conclusion}
\label{sec:conclusion}

This paper defines a bounded, chip-oriented architecture for a one-way implant-to-wearable IR-UWB uplink.
At the implant, a neural-data and packet path independently encodes Base and Enhancement lanes, maps them to \ModeZero{} PWM or \ModeOne{} PWM plus DBPSK, and connects them to an attributed free-running RO/selected-phase edge-combining transmitter, PA, and antenna.
The underlying RO and pulse-generation circuit are prior work by Lei \emph{et al.}; the contribution here is their explicit placement within the end-to-end partition and protocol, not a new transmitter circuit.
Across the modeled package--antenna--tissue channel, the wearable AFE places acquisition, timing support, and Base pulse-width demapping in one envelope branch and uses one gated burst-I/Q chain for optional differential-phase Enhancement data.
The digital baseband and host boundary complete the receive path.
The deterministic protocol chain makes the Header, FEC, interleaving, CRC, AEAD, M0/M1 mapping, differential resets, and independent lane verification correspond across the two endpoints.
\WidthTrain{} supplies packet-local short/long templates and soft width LLRs, while a frozen quality gate selects conservative fallback behavior when those templates are not trustworthy.
Base-only service is reported as a separate, verified degradation outcome rather than complete \ModeOne{} delivery.

The frozen 80{,}000-packet synthetic campaign shows that online packet-local
templates improve average soft-information quality and strict delivery over
scalar and fixed-template baselines, but the gain is not sufficient for a
reliable product claim.
At nominal 18\,dB, same-distribution strict Complete PSR reaches 88.1\% for
M0/W1 and 75.8\% for M1/W1, then falls to 66.1\% and 41.6\% under the
predeclared shifted stress population.
The tested quality-gated fallback improves failure semantics but not average
strict delivery.
The paired 256/512-sample check finds phase-path sensitivity near the
waterfall, so the frozen 256-sample DBPSK result is not presented as
numerically converged.
A separate benign single-tap sensitivity study shows that fixed two-copy soft
combining can move the 12\,dB point estimate above 85\%, but it doubles
scheduled energy and latency and still does not establish medical-grade PER.
At 18\,dB the normal single-copy profile already observes 500/500 successes in
that benign condition and retains substantially higher goodput, so blind
repetition is a critical-packet mode rather than a default stream mode.

The immediate research needs are a measured implant-package/tissue/wearable
channel and front-end reference plane, circuit implementation of the wearable
receiver, validation of the TX integration and power-gating assumptions, and
receiver algorithms that reduce the observed acquisition, Header, and payload
losses.
The architecture remains a one-way uplink, not a bidirectional transceiver
product; reference-assisted TX calibration, an implant receiver, and an
external downlink are outside scope.
The conclusions remain limited to simulation and do not imply fabricated
silicon, measured biological-channel, clinical, or regulatory validation.

\bibliographystyle{IEEEtran}
\bibliography{references}

\end{document}